\documentclass[twocolumn,showpacs,preprintnumbers,amsmath,amssymb]{revtex4}

\usepackage{graphicx}
\usepackage{dcolumn}
\usepackage{bm}

\newcommand{\cc}{c \bar c}
\newcommand{\ups}{\Upsilon}
\newcommand{\jp}{J/\Psi}
\newcommand{\ee}{e^+ e^-}

\begin{document}

\preprint{SDU-HEP200603}

\title{Remarks on CLEO  New Measurements for $\ups (1S)$ Decays to
       Charmonium Final States and Investigations on Associate
       Strange Particle Enhancement in $\ups \to \jp +X$}

%

\author{Wei Han}
\affiliation{Department of Physics, Shandong University,
Jinan,  250100, P.R. China}
\email{hanwei@mail.sdu.edu.cn}

\author{Shi-Yuan Li}
 \affiliation{Institute of Particle Physics, Huazhong Normal
University,  Wuhan, 430079, P.R. China}
%
 \email{lishy@sdu.edu.cn}
  \altaffiliation[on leave from]
{~{\it Department of Physics, Shandong University,
Jinan,  250100, P.R. China}}



\date{\today}

\begin{abstract}
 The recent  measurements by CLEO 
for  the 
  inclusive   $ \jp $ and $\Psi(2S)$ production in $\ups (1S)$  decay
and our previous calculation
are analyzed.
The $\jp$ momentum spectrum and the production
ratio of  $\Psi (2S)$ versus  $\jp $ 
favour  $\ups \to \jp (\Psi (2S)) + \cc g$ as the dominant
contribution. 
We point out that the differences between the experimental data and our 
previous  results  are mainly
originated from the  setting of   the  parameter charm quark mass.
 Encouraged by this result, we investigate    the associate
  strange particle enhancement  as a probe
for the open charm particles in 
   $\ups \to \jp (\Psi (2S)) + \cc g$.
\end{abstract}

\pacs{13.25.Gv, 13.87.Fh, 12.38.-t}
\keywords{$\jp$ momentum spectrum, production ratio,  
strange enhancement}
\maketitle

Besides the important r\^{o}le in the electroweak and CP violation 
research, heavy
quark (b or c)  physics
 is also the  traditional arena for Quantum
Chromodynamics (QCD), both in
perturbative (P) and non-perturbative (NP) aspects. 
The production and annihilation  of heavy 
quarks via strong interactions 
can be calculated by PQCD for their large masses naturally set the hard
scales, while the structures of heavy hadrons
 are ruled by NPQCD. 
Since  heavy quarks can hardly be produced in NPQCD
phase,  
one can compare the quantities 
of  them  calculated by PQCD
with those of the corresponding heavy hadrons
to study the NPQCD effects.
Hence, heavy quarks  act  as  bridges 
linking  the P and NP phases of QCD. 
Nevertheless, such a distinct
character requires a careful investigation on both PQCD and NPQCD
phases for a certain  process, as one has to do in the 
 study of  quarkonium production.

Non-Relativistic QCD (NRQCD) \cite{Bodwin:1994jh}
provides a factorization framework \cite{sterman} to calculate the heavy 
quarkonium production and decay. 
For example,  the inclusive production of charmonium $H_c$ in the  
$ p \bar p $ collision  can be written as 
\begin{equation}
\label{NRQCDFAC}
d \sigma(p \bar p \to H(P)_c +X)=\sum_i d \hat \sigma (p\bar p \to \cc 
(P)_i +X)  <O^{H_c}_i>,
\end{equation}
which is a sum of various contributions.
Each contribution  is the  product of two factors:  One is 
 the   cross section   of  the 
quark pair production in a definite (colour, angular momentum, etc.) state, 
the other is  the NRQCD
matrix elements 
to describe the transition probability of the quark pair
to the relevant quarkonium.
This formula clearly implies  that 
one has to take into account all the competing
contributions  in calculating the  cross section measured by experiments.
For the inclusive charmonia production
 in $p\bar p$ interaction at Tevatron \cite{cdf1},
 the  colour-octet process \cite{color_octet_mech}  
  seems necessary to explain
 the total cross sections as well as
the transverse momentum spectra 
(But it is not clear yet  why the 
polarization of $J/\Psi$ not properly described \cite{cdf2}),
because only the cross section of all  the 
colour-singlet processes summed up  
is of one order lower than the experimental data. This is one example
that teaches us to include all kinds of contributions, but 
does not mean colour-octet process(es)  always the dominant. 
 Until now, the values of 
colour-octet matrix elements can only be extracted via 
fitting the data but are generally believed to be smaller 
than the colour-singlet ones, based upon the 
NRQCD velocity counting rule.  
Furthermore, when the same matrix elements determined 
at CDF are
applied to photoproduction of $\jp$ at HERA, the color-octet
contribution is about a factor of ten too large \cite{h1}.
This indicates that these matrix elements may be even smaller. 
So, for a concrete process,
 one can not judge \`{a} priori which kind of process is 
the main contribution 
 and  can  not 
truncate the hard part   
($d \hat \sigma$ in Equation (\ref{NRQCDFAC})) 
to a certain order of $\alpha_s$, 
without distinction for colour-singlet or
colour-octet partonic processes. The reason is that the 
suppression from $\alpha_s$ for higher order colour-singlet  partonic
process(es) 
can be compensated by the larger  colour-singlet  matrix 
element(s) 
 (The colour-singlet matrix elements of vector mesons 
can be determined unambiguously from their leptonic decays).
So higher order (in $\alpha_s$) colour-singlet 
process could also contribute
significantly. 
This is the key point \cite{Li:1999ar} to understand the 
inclusive $\jp$ production data
 in $\ups$ decay \cite{Briere:2004ug, Fulton:1988ug}.

The strong decay of $\ups$ is dominantly via 3 gluons, because of 
charge parity conservation. 
So the straightforward partonic process
 for the charmonium
 production  is $\ups \to ggg* \to gg c \bar c$ \cite{cheung_1s},
which is of  lowest order  PQCD to create charm quark pair. 
Here the 
charm quark pair  is produced via the virtual gluon splitting 
and is inherently colour-octet.
When the  colour-octet matrix element extracted from Tevatron 
is applied, one can get the  branching ratio comparable to the 
experimental data
\cite{Fulton:1988ug}. However, because one of the 
real gluon is preferred to be soft, the $\jp$ momentum spectrum
peaks at the maximum value, like a 2-body final state
configuration.
This was not consistent with the old CLEO measurements \cite{Fulton:1988ug},
  but  the error bar was too large
to draw a final conclusion, then. Another colour-octet process which is also 
of the  2-body final state configuration was  suggested \cite{napsuciale}, 
ignorant of the  $\jp$ momentum spectrum, too.
Recently, the CLEO collaboration make a new measurement 
for the $\ups(1S)$ decays to charmonium final state \cite{Briere:2004ug}.
The data sample is 35 times larger. The more precise measurement 
confirm a rather soft spectrum of the $\jp$ in the $\ups \to \jp +X$ process, 
which is consistent with the colour-singlet process prediction
\cite{Li:1999ar} both in the branching ratio as well as the $\jp$ spectrum 
but clearly 
conflict with the colour-octet processes in $\jp$ momentum spectrum.

As is discussed in \cite{Li:1999ar} and in above, because the much
larger colour-singlet matrix element(s) can compensate the suppression from
powers of $\alpha_s$, one should not ignore the colour-singlet process(es)
if the suppression from powers of $\alpha_s$ is not too large
(e.g. here,  only one extra $\alpha_s$).
 Hence we calculated the 
contribution of $\ups \to \jp + c \bar c  g$ to the 
inclusive $\jp$ production of $\ups$ decay \cite{Li:1999ar}.
This is dominant of colour-singlet process and can be treated within 
 the traditional non-relativistic wave function approach 
\cite{khun, Chang:1979nn, baier}. For this process,  
it is very clear that the
 more final state particles, especially the open
charm hadrons from  the unbound charm quark pair
 make the $\jp$ spectrum much more soft.  
 That the dominant contribution of this channel is via 
colour-singlet process
requires that the bound $c$ and $\bar c $ should come from different 
virtual gluons and in nearly the same momentum. This  configuration 
in phase space eliminates the
 2-body like configuration that the real  gluon is soft 
and the 2 virtual gluon splitting 
into 2 back to back  $c \bar c$ pair with minimum invariant mass.
So the momentum spectrum of $\jp$ can not peak at the largest value.
Though the qualitative analysis is  perfectly consistent with data, 
the new data \cite{Briere:2004ug} 
 and our calculation \cite{Li:1999ar} does not coincide completely.
One may be interested whether there are other dynamical reasons  ignored.
Hence in the following, we first  
clarify  that the ``inconsistency'' 
between the data and our calculation
 comes from the setting of the parameter charm mass
in numerical calculation.
We 
 emphasize on  the  $\jp$ momentum spectrum 
and the  production
ratio of $\Psi (2S)$ versus  $\jp $.
 Furthermore,  we investigate  the associate 
strange particle enhancement as a way 
 to  probe the charm quark accompanying with $\jp$.

~~~~~~~~~~

{\large \it The reason for the ``harder'' spectrum of our 
previous calculation  than that of the  experiment}

The CLEO collaboration claims that the momentum spectrum is ``closer
to, although softer than'' our  colour-singlet prediction
\cite{Briere:2004ug}. However,
this is just because we did not
take into account the open charm threshold effect
in the partonic level calculation. In our calculation,
we  let the charm quark mass equals to half of $\jp$ mass. This leads to
the unphysical region when the $\cc g$ cluster mass smaller than $2 m_D$.
If taking into account the threshold  effect, 
the peak in our calculation should 
move to the left  about $0.1$ unit of $x$ (scaled momentum). This means 
 the position of the peak of our calculated $\jp$ momentum
spectrum (see, FIG. 3 of \cite{Li:1999ar}  or FIG. 10 of \cite{Briere:2004ug})
should move to around  $x=0.4$ rather than $x=0.5$. 
 Considering that the width of the bins of the experimental data
is 0.2 unit, the prediction of the colour-singlet model
 is consistent rather than softer than 
the data.

~~~~~~~~~

{\large \it The ``wrong'' result of our calculation of $\Psi (2S)$
production rate}

If we only consider the production rate of $\Psi (2S)$ itself, 
it may be 
natural to choose  $m_c=M_{\Psi (2S)}/2$ and use the  $\Psi (2S) \to \ee$
decay width to extract the wave function
(In fact sometimes one needs to tune the quark mass to get the 
proper  absolute value of the cross section). This is what we did 
in our previous calculation. 
In  \cite{Li:1999ar},
when calculating the partonic cross section of charm pair creation
for $\jp$ production, 
we use $m_c=M_{\jp}/2$; while 
calculating  that for the $\Psi(2S)$ production, we use $m_c=M_{\Psi(2S)}/2$. 
As can be seen from the Table demonstrating the charm mass dependence,
this leads
to a difference of about 2.5 times. The ratio of the   wave functions  square
and the branching ratio of $\Psi(2S)$ to $\jp$ lead to another factor about
4. This is why in our paper we only predicted  about 10 per cent of $\jp$ are
from the decay of $\Psi (2S)$.
However, if we calculate the production ratio(es) between  
 $\jp$, $\Psi (2S)$, etc.,  and adopt that they 
are different eigenstates of $\cc$ bound state,
 their differences should be only inherent  in the wave functions. The 
$M_{\Psi(2S)}-M_{\jp}$ is explained as 
originated from difference of  binding energies.
Hence  the charm quark mass should 
be  taken as the same. This makes the calculated results 
consistent and simple for the colour-singlet processes. 
In the framework 
of non-relativistic wave function approach \cite{khun, Chang:1979nn, baier},
 one can easily derive, 
for the S-wave particles,
\begin{eqnarray}
\label{nwf}
 \frac{{\cal B}(\ups \to \Psi (2S)+\cc g)} {{\cal B}
(\ups \to \jp +\cc g)} &
=&\frac{\Gamma(\ups \to \Psi (2S)+\cc g)} {\Gamma
(\ups \to \jp +\cc g)} \nonumber \\
 &=& \frac{|\psi^{\cc}_{2S}(0)|^2}{|\psi^{\cc}_{1S}(0)|^2} \nonumber \\
 &=& \frac{\Gamma (\Psi (2S) \to \ee)} { \Gamma (\jp \to \ee)}.
\end{eqnarray}
In the above equations,
$\psi^{\cc}_{2S}(0)$ and $\psi^{\cc}_{1S}(0)$ are wave functions at origin
for $\Psi (2S)$ and $\jp$, respectively.
To get the relations in Equation (\ref{nwf}), we notice that
$ \Gamma(\ups \to \Psi (nS)+\cc g)$  should be,
to order $O(v^0)$,  
\begin{eqnarray}
\label{app}
~&\frac{1}{2 M_{\ups}} \frac{|\psi^{\cc}_{nS}(0)|^2}{m_c}
\int_{2m_c} \sqrt{E^2-4m^2_c}~ dE \int dR' \nonumber \\
\times &\overline{|M(\ups \to c(P/2) \bar c(P/2) \cc g)|^2}, 
\end{eqnarray}
where  $dR'$ represents other integral variables than 
the $\Psi(nS)$ energy $E$ for the $\Psi(nS)\cc g$ phase space element. 
P is the four momentum of $\Psi(nS)$.
$M(\ups \to c(P/2) \bar c(P/2) \cc g)$ is the invariant amplitude 
for $\ups \to c(P/2) \bar c(P/2) \cc g)$ process.
In this formula, when taking  the same charm quark mass,
the only difference for different S-wave bound states is the wave function
square.
For the leptonic decay width of S-wave vector charmonia, 
a very similar formula and hence conclusion  can be obtained. Only that 
one should take the energy of the charmonia at rest frame to be
$2m_c$, neglecting their mass differences.

Using the values of 2002 data book \cite{pdg}, 
which is also  used
by the experimental group \cite{Briere:2004ug}, 
one can calculate the ratio of the $\ee$ decay widths (central value) between 
$\Psi (2S)$ and $\jp$. It is striking to be 0.416
(the to-date values
 give  0.45), which is very near the 
 central value of the data 0.41 \cite{Briere:2004ug}. 
However, the colour-octet process can not simply derive 
such a relation with the $\ee$ decay width\footnote{%
The ratio of the corresponding colour-octet matrix elements of 
$\Psi(2S)$ and $\jp$ \cite{barpn} 
 is around 0.3, which is lower than but within the error of the data.}

Unfortunately, we have not got the analytical formula for
the partonic process $\jp \to \cc \cc g$, so it is difficult
to evaluate 
the P wave particles which is also measured by CLEO.
The calculation needs the derivatives of the partonic 
amplitude. 
Encouraged from the S-wave productions, 
we leave the calculation of P-wave particles as an interesting 
separate work to do.

From the above discussions, it is clear not that 
NRQCD leads to the prediction that the colour-octet 
processes dominate  in $\ups \to \jp +X$. It is just 
the NRQCD formulae require a
 global consideration of all possible partonic processes
 to explain the data.
Frankly, 
there are  many uncertain parameters 
like quark mass, $\alpha_s$ (scale dependence is significant 
in the lowest order calculation), etc.
So only quantities like the relative production ratio 
 and shape of spectrum which are not sensitive to the absolute values of 
these parameters are relevant to justify  
the partonic processes. Only to get a cross section or branching ratio 
comparable with data sometimes seems tricky.

~~~~~~~~~~

{\large \it Associate strange particle enhancement}
The above arguments confirm the significance of the colour-singlet
process. 
As has been pointed out by  \cite{Li:1999ar}
 and the experimental group \cite{Briere:2004ug}, 
a very important probe 
for the partonic process is to measure the content of the 
cluster produced associated with the $\jp$ in the $\ups$ decay,
which is theoretical-framework-independent.
It is  interesting to notice  that 
in $\ee $ continuum process at 
nearly the same energy,   BELLE Collaboration 
\cite{bw} reported that about 60  per cent events of 
$\ee \to \jp +X$ is from $\ee \to \jp +\cc$.
This is a challenge for the theory,
and also  suggests the importance to measure the 
associate  particles in 
$\ups \to \jp +X$.
However, the data 
sample seems still not enough. 
According to  the experiment \cite{Briere:2004ug, private}, 
the $\ee$ and $\mu^+ \mu^-$ pairs  are  used to 
identify and  reconstruct  the charmonium.  
Only less than $10^3$  $\jp$ events are  collected. Even if  these
events are all $\ups \to \jp + \cc g $, 
and employing  the standard method to reconstruct all the 
charm pseudo-scalar and vector mesons that the associate charm quarks
fragment to
one can  only 
expect to get 
less than  10 per cent events. 
On the other hand,
it is  not efficient to reconstruct all the important charm mesons.
If only use the  $D^0$ ($\to K \pi$) channel,  
the number of events to be found 
could only be around 9 \cite{private}.
%
Then the  statistical error may not allow  a definite conclusion.

However, a charm quark generally  decays into strange
quark, hence the   
strange particles can manifest the ever-production of charm. 
So the inclusive measurement of kaon may be a good probe \cite{private}.
Besides  charm decay, the strange quark can 
be created from the vacuum during the hadronization
process. Hence  it is not the absolute production 
rate of strange particles
but their 
enhancement relative to other cases is the very signal for charm.
%
For the processes we are going to investigate, 
the colour-octet models predict 
a flavour-singlet cluster (one or two colour-octet gluon(s)) associate  
with  $\jp$, while the colour-singlet model predicts 
the cluster $c \bar c g$,
which lead to strange flavour via weak decay.
To find a measurable quantity, we study the 
ratio $\frac{<K>}{<\pi>}$  
($<K>$ and $<\pi>$ are average multiplicity of $K^{\pm}$ and $\pi^{\pm}$,
 respectively) of the cluster accompanying with the $\jp$.
The cluster  $c \bar c g$
is colour-singlet and we can straightforwardly
employ JETSET \cite{jetset} to simulate its hadronization \footnote{We 
use the   subroutine lu3ent.
The invariant mass of this cluster is set between 
$M_{2D}$ to $M_{\ups}-M_{\jp}$,
with the weight calculated by the same programme to calculate the $\jp$
spectrum.}. We can expect the 
ratio  is similar as those of $\ee \to c \bar c \to h's$
at the same energy (see Table 1).
 On the other hand, it is not very practical 
to hadronize the {\bf colour-octet}  2-gluon-cluster with JETSET
for the colour-octet process $\ups \to \jp + gg$.  
Another ``technical'' problem is that 
in the partonic level calculation,
this  cluster dominantly have a vanishing  invariant mass. 
For providing a reference to the charm enhancement,
we give the result of a cluster of 
$gg$  which is in colour-singlet 
(its invariant mass is set to be the peak value 
of   the $\cc g$ cluster mass). 
What is  in common for  the 2-gluon cluster
in colour-octet and colour-singlet 
is that they are both flavour-singlet.
We find that the $\frac{<K>}{<\pi>}$ value is 
not sensitive to the invariant 
mass for the colour-singlet $gg$ cluster.
At the same  time, using the JETSET subroutine (lu1ent), 
one can investigate the 
``independent'' fragmentation of one gluon
in the colour-octet process $\ups \to \jp +g$.   
 The obtained  $\frac{<K>}{<\pi>}$  
is insensitive
to its energy. We find that the   result is near that of colour-singlet 
$gg$ cluster (Table 1).   
For  2-gluon system, if we 
do not care it is in colour-singlet or octet, its $\frac{<K>}{<\pi>}$
value can be obtained from independent fragmentation of each gluon,
which is very near that  of the colour-singlet 2-gluon cluster
from the above discussion. This
suggests that the colour-octet 2-gluon cluster should give 
$\frac{<K>}{<\pi>}$ not far beyond that of the colour-singlet cluster.
Besides the above, 
for a comprehensive comparison,
the results of a  flavour-singlet virtual photon ($M<2m_b$)
are also given
as reference,
but we should understand that the electromagnetic interaction breaks the 
flavour  
SU(4) invariance (the initial quark effects).
The result is also shown in Table 1.

In Table 1, we  give the ``errors'' of the calculated results.
Their definition is based upon the following consideration:
The number of collected events by the CLEO Collaboration is only of the order  
$10^3$. So we should estimate the statistical error, or  
the fluctuation of some $10^3$ event samples.
Here we use $rot$  to represent the ``real'' value of 
$\frac{<K>}{<\pi>}$, which
can be estimated by a sample generated by JETSET 
with event number large enough ($\to \infty$).
At the same time, we generate many $10^3$ event samples,   
then we calculate 
\begin{equation}
\chi=\sqrt{\lim_{N\to \infty} \sum_{j=1}^N \frac{(r_j-rot)^2}{N}}, ~
with ~ r_j=\frac{<K>_j}{<\pi>_j}.
\end{equation}
The subscript j represents  each $10^3$ event sample.
In Table 1, the $\frac{<K>}{<\pi>}$ values are given in the form
$rot\pm \chi$. 
for the processes we discuss above. 
 We can conclude that   the fluctuations are small and 
will not smear the difference, 
so that 
 the strange particle enhancement of the 
$\cc g$ cluster 
is very significant, comparing with the flavour-singlet g (gg)
as well as the virtual photon.

\begin{table}
\begin{tabular}{|l|c|}  \hline
processes &$\frac{<K>}{<\pi>}$ \\ \hline \hline
{\em $\cc g \to h's $}&0.243$\pm$0.009 \\ \hline
{\em $\ee \to \cc \to h's$}&0.240$\pm$0.008 \\ \hline
{\em $gg \to h's$}& 0.120$\pm$0.007 \\ \hline
{\em $g \to h's$}&0.113$\pm$0.007 \\ \hline
{\em $\gamma^* \to h's$} &0.176$\pm$0.008 \\ \hline \hline
\end{tabular}
\caption{The production ratio between $K$ and $\pi$ in various processes.}
\label{corr}
\end{table}

In summary, we point out that the recent measurements for $\ups$
decay into charmonium final states by CLEO Collaboration agree well
with the $\ups \to \jp +\cc g$ calculation, provided  that the value of 
charm quark mass is 
properly set. 
This indicates that the colour-singlet process 
could be the dominant  for
the inclusive $\jp$ or $\Psi(2S)$ production in $\ups$ decay.
Furthermore, by  investigating the 
hadronization of the parton(s) produced accompanying with 
the $\jp$ in $\ups$ decay, we found that  
the $\cc g$ cluster in 
colour-singlet (but not flavour-singlet)  will lead to 
significant strange particle enhancement
comparing to the colour-octet (but flavour-singlet) gluon(s).
Thus it is a good signal to probe the process
$\ups \to \jp + \cc g$.

~~~~~~~~

We thank Prof. S. Blusk for  informing  us the 
experimental results and
 following discussions  that push forward this investigation.
We also thank Dr. Yu-Kun Song for the discussions on gluon fragmentation 
function.

\appendix

\section{The distribution of  $\frac{<K>}{<\pi>}$}
In the following figure, we demonstrate the  distribution of the
$\frac{<K>}{<\pi>}$, to show that the dispersions are really very small.

\begin{figure}
\includegraphics[scale=0.40]{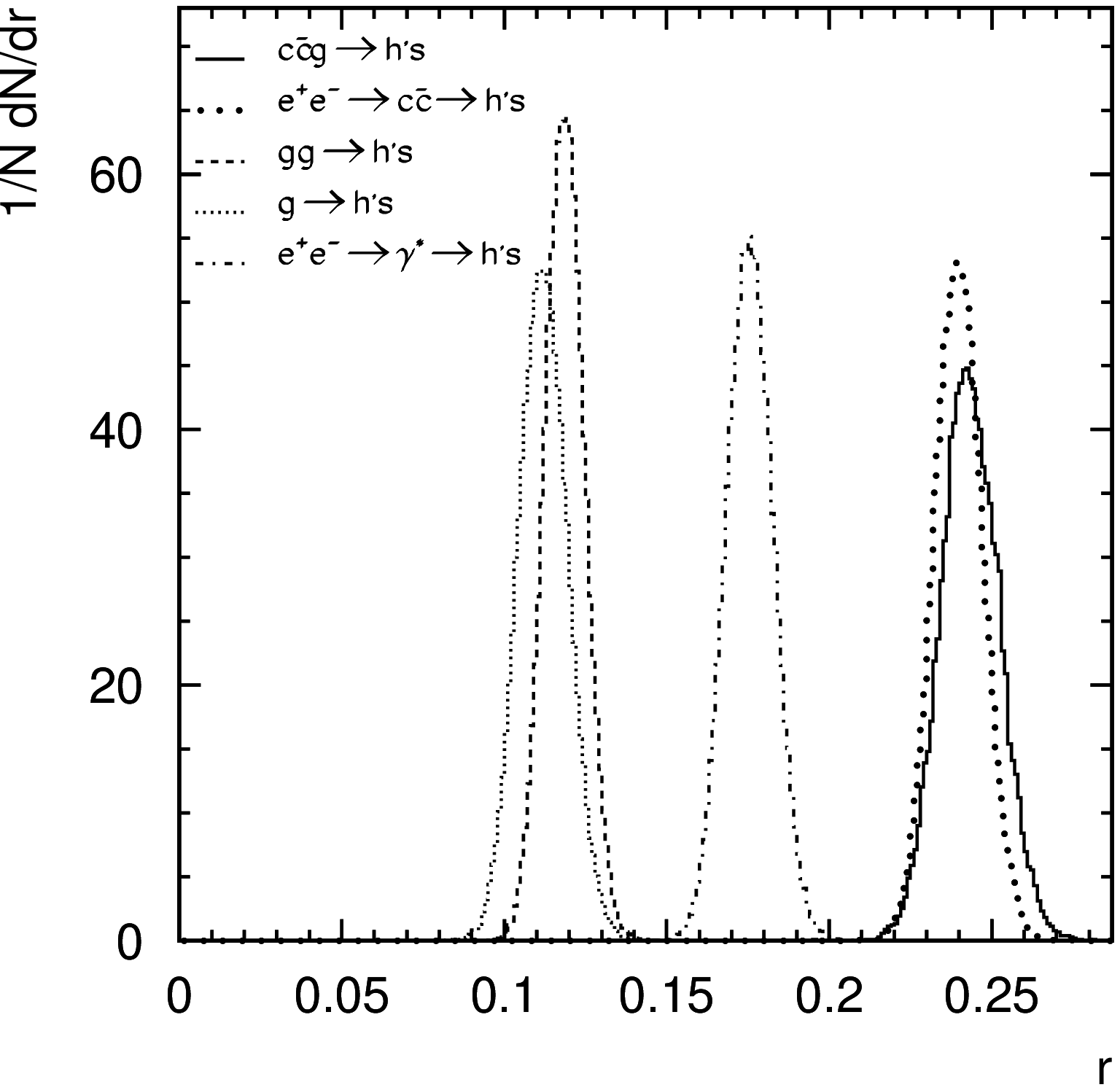}
\caption{Here $r=\frac{<K>}{<\pi>}$}
\end{figure}

\end{document}